\documentstyle[12pt]{article}

\def\PRL #1 #2 #3{{\sl Phys. Rev. Lett.} {\bf#1} (#2) #3}
\def\PRB #1 #2 #3{{\sl Phys. Rev.} {\bf B#1} (#2) #3}
\def\NPBFS #1 #2 #3 #4{{\sl Nucl. Phys.} {\bf B#2} [FS#1] (#3) #4}
\def\CMP #1 #2 #3{{\sl Commun. Math. Phys.} {\bf #1} (#2) #3}
\def\PRD #1 #2 #3{{\sl Phys. Rev.} {\bf D#1} (#2) #3}
\def\PLA #1 #2 #3{{\sl Phys. Lett.} {\bf #1A} (#2) #3}
\def\PLB #1 #2 #3{{\sl Phys. Lett.} {\bf #1B} (#2) #3}
\def\JMP #1 #2 #3{{\sl J. Math. Phys.} {\bf #1} (#2) #3}
\def\PTP #1 #2 #3{{\sl Prog. Theor. Phys.} {\bf #1} (#2) #3}
\def\SPTP #1 #2 #3{{\sl Suppl. Prog. Theor. Phys.} {\bf #1} (#2) #3}
\def\AoP #1 #2 #3{{\sl Ann. of Phys.} {\bf #1} (#2) #3}
\def\PNAS #1 #2 #3{{\sl Proc. Natl. Acad. Sci. USA} {\bf #1} (#2) #3}
\def\RMP #1 #2 #3{{\sl Rev. Mod. Phys.} {\bf #1} (#2) #3}
\def\PR #1 #2 #3{{\sl Phys. Reports} {\bf #1} (#2) #3}
\def\AoM #1 #2 #3{{\sl Ann. of Math.} {\bf #1} (#2) #3}
\def\UMN #1 #2 #3{{\sl Usp. Mat. Nauk} {\bf #1} (#2) #3}
\def\FAP #1 #2 #3{{\sl Funkt. Anal. Prilozheniya} {\bf #1} (#2) #3}
\def\FAaIA #1 #2 #3{{\sl Functional Analysis and Its Application} {\bf
#1} (#2) #3}
\def\BAMS #1 #2 #3{{\sl Bull. Am. Math. Soc.} {\bf #1} (#2)
#3}
\def\TAMS #1 #2 #3{{\sl Trans. Am. Math. Soc.} {\bf #1} (#2) #3}
\def\InvM #1 #2 #3{{\sl Invent. Math.} {\bf #1} (#2) #3}
\def\LMP #1 #2 #3{{\sl Letters in Math. Phys.} {\bf #1} (#2) #3}
\def\IJMPA #1 #2 #3{{\sl Int. J. Mod. Phys.} {\bf A#1} (#2) #3}
\def\AdM #1 #2 #3{{\sl Advances in Math.} {\bf #1} (#2) #3}
\def\RMaP #1 #2 #3{{\sl Reports on Math. Phys.} {\bf #1} (#2) #3}
\def\IJM #1 #2 #3{{\sl Ill. J. Math.} {\bf #1} (#2) #3}
\def\APP #1 #2 #3{{\sl Acta Phys. Polon.} {\bf #1} (#2) #3}
\def\TMP #1 #2 #3{{\sl Theor. Mat. Phys.} {\bf #1} (#2) #3}
\def\JPA #1 #2 #3{{\sl J. Physics} {\bf A#1} (#2) #3}
\def\JSM #1 #2 #3{{\sl J. Soviet Math.} {\bf #1} (#2) #3}
\def\MPLA #1 #2 #3{{\sl Mod. Phys. Lett.} {\bf A#1} (#2) #3}
\def\JETP #1 #2 #3{{\sl Sov. Phys. JETP} {\bf #1} (#2) #3}
\def\JETPL #1 #2 #3{{\sl  Sov. Phys. JETP Lett.} {\bf #1} (#2) #3}
\def\PHSA #1 #2 #3{{\sl Physica} {\bf A#1} (#2) #3}
\def\CQG #1 #2 #3{{\sl Class. Quantum Grav.} {\bf #1} (#2) #3}
\def\SJNP #1 #2 #3{{\sl Sov. J. Nucl. Phys. (Yadern.Fiz.)} {\bf #1} (#2) #3}

\def\be{\begin{equation}}
\def\ee{\end{equation}}

\input{tcilatex}

\begin{document}

\newcounter{oldequation}



\vspace{0cm}

\begin{center}
{\large {\bf Quasi-one-dimensional charge density wave in electromagnetic
field arbitrarily oriented to conducting chains: generalized Fr\"ohlich
relations}}

\vspace{2cm} Alexander S. Rozhavsky$^{(1,2,a)}$, Yurij V. Pershin$^{(1,2)}$
and Igor A. Romanovsky$^{(1,3)}$

\vspace{0.5cm} $^1${\it B.I.Verkin Institute for Low Temperature Physics and
Engineering, \\[0pt]
47 Lenin Avenue 310164 Kharkov, Ukraine \\[0pt]
(permanent address)}

\vspace{0.5cm} $^2${\it Grenoble High Magnetic Field Laboratory, \\[0pt]
Max-Planck-Institut f\"ur Festk\"orperforschung and CNRS, \\[0pt]
BP 166, F-38042, Grenoble, Cedex 9, France}

\vspace{0.5cm} $^3${\it Physical Department, Kharkov State University, \\[0pt%
]
4 Svobody Square 310077 Kharkov, Ukraine}

\vspace{2cm} 
{\bf Abstract}

\vspace{1cm}
\end{center}

We derive equations for the collective CDW-current transverse conducting
chains in a quasi-one-dimensional CDW-conductor. Generalized Fr\"ohlich
relations between the transverse currents and phase gradients are due to the
polarization corrections to the $1+1$ chiral anomaly Lagrangean. The CDW
Hall constant $R_{CDW}$ is calculated, $R_{CDW}\sim T^2_C/I_{CDW}$, where $%
T_C$ is the critical temperature of the Peierls transition, and $I_{CDW}$ is
the nonlinear CDW current in the direction parallel to the conducting chains.

\vspace{2cm} PACS numbers: 71.45Lr, 72.15Nj, 11.30Rd \vspace{1cm}

$^{a)}$ Corresponding author. E-mail: rozhavsky@ilt.kharkov.ua

\renewcommand{\thefootnote}{\arabic{footnote}} \setcounter{footnote}0
\newpage

\renewcommand{\thefootnote}{\arabic{footnote}} \setcounter{footnote}0
The Quasi-One-Dimensional (Q1D) Charge Density Wave (CDW) conductors such as 
$NbSe_3$, $TaS_3$, $K_{0,3}MoO_3$ etc. (see for instance, Ref.\cite{review})
are characterized by a strongly anisotropic quasiparticle spectra. As a
result, their unusual transport properties are mostly pronounced in the
direction parallel to the conducting chains. The theoretical studies of the
CDW-electrodynamics were mainly focused on the one-dimensional aspect. The
response of a 1D CDW is described by the well-known Fr\"ohlich relations 
\cite{review}:

\begin{equation}
j_{x}=-\frac{e}{\pi }\frac{\partial \varphi }{\partial t}n_{f}
\label{frolrho} \\
\end{equation}
\[
\rho =\frac{e}{\pi }\frac{\partial \varphi }{\partial x}n_{f}
\]
where $j_{x}$ is the collective current density along the direction of the
chains, $\rho $ is the charge density fluctuation, $n_{f}$ is the density of
chains and $\varphi $ is the CDW-variable, the phase of the Peierls-Fr\"{o}%
hlich order parameter $\Delta exp(i\varphi )$. The amplitude $\Delta $ is
equal to the gap width in the single-particle spectrum.

Experimental investigations of the transverse CDW transport, and in
particular of the Hall effect \cite{l2,l3}, have demonstrated that the
3D-effects are also strongly affected by the nonlinear CDW transport current 
$I_{CDW}$ directed along the chains. It was found in Ref. \cite{l3} that the
Hall constant $R_{CDW}$ is proportional to $I_{CDW}^{-1}$ and sharply
decreases in electric fields above the threshold $E_{T}$. This fact is not
yet explained within the microscopic theory. To describe properly the 3D
electrodynamics of CDW, we need the generalized Fr\"{o}hlich equations for
all the three components of the collective current $\vec{I}_{CDW}$. This
problem is solved in the present paper, and the dependence of $R_{CDW}$ on $%
I_{CDW}$ is obtained.

The description of a Q1D CDW is based on the electron-lattice Hamiltonian:

\[
\hat{H}=-\frac{\hbar^2}{2M}\sum_{\vec n_{\nu}}\frac{\partial^2} {\partial {%
u^2_{\vec n_{\nu }}}}+\sum_{\vec n_{\nu}}K_{\nu} \left( u_{\vec n_
{\nu}}-u_{\vec n_{\nu}+ \vec1} \right)^2+ 
\]
\begin{equation}  \label{letham}
\sum_{\vec n_{\nu}}\left(t_{\nu}+t_{1\nu}\left( u_{\vec n_{\nu}}-u_{\vec
n_{\nu}+ \vec1} \right)\right) \left(a^+_{\vec n_{\nu},s}a_{\vec n_{\nu}+
\vec 1, s}+h.c.\right)
\end{equation}
where $u_{\vec n}$ is the lattice displacement, $\vec n$ numerates lattice
sites, $\nu =x,y,z$ , $M$ is the ion mass, $K_\nu$ are the lattice
elasticity constants, $t_{\nu}$ is the electron hopping integral, $t_1$ is
the electron-lattice coupling, $a^+_{\vec ns}$ and $a_{\vec ns}$ are the
creation and annihilation operators of an electron with spin $s$ at the site 
$\vec n$.

The Q1D approximation is formulated as follows:

\[
u_{n_x}=u_0cos(2k_Fn_xb_x+\varphi) 
\]
\begin{equation}  \label{approxim}
a_{\vec ns}=\Psi_{Rs}exp(ik_Fn_xb_x)+\Psi_{Ls}exp(-ik_Fn_xb_x)
\end{equation}
\[
u_{\vec n_{y,z}}-u_{\vec n+\vec 1_{y,z}}\simeq \frac{\partial u_{\vec n}}{%
\partial y(z)}b_{y,z} 
\]
\[
t_{1y,z}=0 
\]
For our purposes, it is convenient to formulate the theory in the Lagrange
formalism.

After the standard redesignations (see e.g. \cite{l4}): 
\[
b_{x}t_{0}=\hbar V_{F},~~2t_{1x}u_{0}/b_{x}=\Delta
,~~K_{x}b_{x}sin^{2}(b_{x}k_{F})/16t_{1x}^{2}=\frac{1}{g^{2}},
\]
the Peierls-Fr\"{o}hlich Lagrangean takes the form ($\hbar =c=1$): 
\[
{\cal L}=\varphi \left( \frac{1}{\alpha ^{2}V_{F}}\partial _{t}^{2}-\frac{%
U_{y}^{2}}{V_{F}}\partial _{y}^{2}-\frac{U_{z}^{2}}{V_{F}}\partial
_{z}^{2}\right) \varphi -\frac{\Delta ^{2}}{2g^{2}}+\bar{\psi}\{i\sigma
_{2}(\partial _{t}+ie\Phi +it_{y}ch(b_{y}(\partial _{y}-ieA_{y}))+
\]
\begin{equation}
it_{z}ch(b_{z}(\partial _{z}-ieA_{z})))-\sigma _{1}V_{F}(\partial
_{x}-ieA_{x})-\Delta exp(-i\sigma _{3}\varphi )\}\psi ={\cal {L}}^{ph}+\bar{%
\psi}L\psi   \label{lagr}
\end{equation}
where the electromagnetic field is introduced in a gauge-invariant form.
Here $\alpha ^{2}=\frac{\Delta }{V_{F}^{2}M}<<1$ is the adiabatic parameter
of the Peierls-Fr\"{o}hlich theory \cite{l4}, $U_{y,z}$ are the transverse
phonon velocities, $\Phi $ and $A_{\nu }$ are the scalar and vector
potentials, $\psi ^{+}=(\psi _{R}^{+},\psi _{L}^{+})$, $\bar{\psi}=\psi
^{+}\sigma _{2}$ and $\sigma _{\nu }$ - are the Pauli matrices. Note that
the approximation (\ref{approxim}) which leads to Eq. (\ref{lagr}) is well
defined only when $t_{y,z}<<\Delta $.

The CDW currents are: 
\begin{equation}
j_{\nu }=c\frac{\delta F}{\delta A_{\nu }}  \label{defrho} \\
\end{equation}
\[
\rho =-\frac{\delta F}{\delta \Phi }
\]
where $F$ is the free energy: 
\[
F=-T\ln \int D\Delta D\varphi D\bar{\psi}D\psi exp\left( \int\limits_{0}^{%
\frac{1}{T}}d\tau \int d\vec{r}{\cal L}_{E}\right) =
\]
\begin{equation}
-T\ln \int D\Delta D\varphi Dexp\left( \int\limits_{0}^{\frac{1}{T}}d\tau
\int d\vec{r}{\cal L}_{E}^{ph}+Tr~\ln L_{E}\right)   \label{statsum}
\end{equation}
Here ${\cal L}_{E}$ is the Euclidean Lagrangean, $\tau =-it$, $T$ is the
temperature and 
\begin{equation}
Tr\hat{O}=2tr\int d\vec{r}d\tau <\vec{r}\tau |\hat{O}|\vec{r}\tau >~~,
\label{spdef}
\end{equation}
where $tr$ denotes trace over the matrix indices, and the multiplier 2
results from the summation over the electronic spins.

To calculate $Tr~\ln L_{E}$, we perform the chiral rotation: 
\begin{equation}
\psi \rightarrow exp(i\frac{\sigma _{3}}{2}\varphi )\tilde{\psi}~,
\label{rot1}
\end{equation}
\[
\bar{\psi}\rightarrow \bar{\tilde{\psi}}exp(i\frac{\sigma _{3}}{2}\varphi )~.
\]
Then: 
\begin{equation}
exp(Tr~\ln L_{E})=J\{\varphi \}exp(Tr~\ln \tilde{L}_{E})  \label{eq1}
\end{equation}
where $J\{\varphi \}$ is the functional Jacobian and the Lagrangean $\tilde{L%
}_{E}$ is: 
\[
\tilde{L}_{E}=\{-\sigma _{2}[\partial \tau +\frac{V_{F}}{2}\partial
_{x}\varphi +e\Phi +t_{y}ch(b_{y}(\partial
_{y}-ieA_{y}))+t_{z}ch(b_{z}(\partial _{z}-ieA_{z}))]-
\]
\begin{equation}
\sigma _{1}V_{F}[\partial _{x}-\frac{1}{2V_{F}}\partial _{\tau }\varphi
-ieA_{x}-\frac{t_{y}b_{y}}{2V_{F}}\partial _{y}\varphi sh(b_{y}(\partial
_{y}-ieAy))-\frac{t_{z}b_{z}}{2V_{F}}\partial _{z}\varphi sh(b_{z}(\partial
_{z}-ieAz))]-\Delta \},  \label{rotlag}
\end{equation}
at $b_{y}\partial _{y}\varphi ,b_{z}\partial _{z}\varphi <<1$.

We calculate the Jacobian following the Fujikawa scheme \cite{fudjik}.
According to this: 
\begin{equation}
J^{-1}\{\varphi \}=exp\left( -\frac{1}{2}\int d\vec{r}d\tau \varphi Tr\chi
^{\ast }\sigma _{3}\chi \right)  \label{jak}
\end{equation}
where $\chi $ stands for the complete set of the asymptotic eigenfunctions
of the Lagrangean $\tilde{L_{E}}$ taken in the ultraviolet limit, i.e. at
the energies far exceeding $e\phi $, $eA_{x}V_{F}$, $t_{y,z}$, and $\Delta $%
: 
\begin{equation}
\chi =exp\left( i(k_{x}x+\omega \tau +k_{y}y+k_{z}z)\right)  \label{planew}
\end{equation}
We have: 
\[
Tr(\chi ^{\ast }\sigma _{3}\chi )=2\lim_{N\rightarrow \infty }\frac{1}{(2\pi
)^{4}}\int\limits_{-\infty }^{\infty }dk_{x}\int\limits_{-\infty }^{\infty
}d\omega \int\limits_{-\pi /b_{y}}^{\pi /b_{y}}dk_{y}\int\limits_{-\pi
/b_{z}}^{\pi /b_{z}}dk_{z}\chi ^{\ast }exp\left( -\frac{\tilde{L_{E}}\sigma
_{3}\tilde{L_{E}}\sigma _{3}}{N^{2}}\right) \chi = 
\]
\[
\frac{2}{b_{y}b_{z}}\Big\{-\frac{1}{2V_{F}\pi }\partial _{\tau }^{2}\varphi -%
\frac{V_{F}}{2\pi }\partial _{x}^{2}\varphi -\frac{e}{\pi }E_{x}+\frac{%
V_{y}^{2}}{2\pi V_{F}}\partial _{y}^{2}\varphi +\frac{V_{z}^{2}}{2\pi V_{F}}%
\partial _{z}^{2}\varphi \Big\}- 
\]
\[
\frac{eb_{y}^{2}t_{y}}{2V_{F}}<ch(b_{y}(\partial _{y}-ieA_{y}))>\left( E_{y}-%
\frac{1}{2e}\partial _{xy}^{2}\varphi \right) \partial _{y}\varphi
-(y\rightarrow z)-\newline
\]
\begin{equation}
ieb_{y}t_{y}<sh(b_{y}(\partial _{y}-ieA_{y}))>\left( H_{z}+\frac{1}{2V_{F}e}%
\partial _{y\tau }^{2}\varphi \right) -(y\rightarrow z)
\end{equation}
Here 
\begin{equation}
<\hat{f}>=\int dk_{x}dk_{y}dk_{z}d\omega \chi ^{\ast }\hat{f}\chi ,
\end{equation}
and $E_{\nu }$, $H_{\nu }$ are electric and magnetic fields. It is readily
seen, that: 
\begin{equation}
<ch(b_{\nu }(\partial _{\nu }-ieA_{\nu }))>=<sh(b_{\nu }(\partial _{\nu
}-ieA_{\nu }))>=0  \label{fields}
\end{equation}
and the Jacobian takes the form: 
\begin{equation}
J\{\varphi \}=exp\left( -\int d\vec{r}d\tau l\{\varphi \}\right)
\label{jakob}
\end{equation}
where 
\begin{equation}
l\{\varphi \}=\frac{1}{b_{y}b_{z}}\Big\{\frac{1}{4V_{F}\pi }(\partial _{\tau
}\varphi )^{2}+\frac{V_{F}}{4\pi }(\partial _{x}\varphi )^{2}-\frac{V_{y}^{2}%
}{4\pi V_{F}}(\partial _{y}\varphi )^{2}-\frac{V_{z}}{4\pi V_{F}}(\partial
_{z}\varphi )^{2}-\frac{e}{\pi }E_{x}\varphi \}~~,  \label{phasel}
\end{equation}
and $V_{y,z}=b_{y,z}t_{y,z}$.

In a $1D$ CDW $(V_{y}=V_{z}=0)$, the Lagrangean (\ref{phasel}) coincides
with the one for the chiral anomaly in $1+1$ massless quantum field theory
(see e.g. \cite{l6}). The connection between the $1D$ CDW Lagrangean and the
chiral anomaly phenomenon was demonstrated in Refs. \cite{l7,l8}. The Fr\"{o}%
hlich relations (1) evidently follow from the Lagrangean (\ref{phasel}).

So, in a Q1D CDW, the chiral anomaly mechanism gives the wrong sign only in
the transverse dispersion in Eq. (\ref{phasel}). The normal CDW-dispersion
is restored due to the phonon terms in the Lagrangean (\ref{lagr}) \cite{l9}%
. A modification of Fr\"{o}hlich relations arises due to the polarization
terms in the free energy hidden in $Tr~\ln \tilde{L}_{E}$ (Eq. (\ref{eq1})).

In Ref. \cite{l10} the chiral anomaly was studied in a formally similiar
system: the Q1D spin density wave. It was claimed in \cite{l10} that, under
the conditions of a quantum Hall effect, the chiral anomaly produces
additional terms linear in $t_y$ in Eqs.(1). In this paper, we study the
opposite limit of classically "weak" magnetic fields (Eq. (\ref{fields}))
when the terms linear in $t_{y,z}$ turn to zero.

Consider $Tr~\ln \tilde{L}_{E}$ at $|\vec{E}|,|\vec{H}|<<\Delta ^{2}/eV_{F}$
and $\vec{\partial}\varphi <<\Delta /V_{F}$, $\partial _{\tau }\varphi
<<\Delta $. The straightforward perturbation expansion (see e.g. \cite{l4})
gives: 
\[
Tr~\ln \tilde{L}_{E}=\frac{1}{2}Tr~\ln \tilde{L}_{E}\sigma _{3}\tilde{L}%
_{E}\sigma _{3}\cong 
\]
\begin{equation}
\frac{1}{2}\ln \hat{K}_{0}-\frac{1}{2}\int d\vec{r}d\tau \big\{\kappa _{xx}%
\tilde{E}_{x}^{2}+\kappa _{yy}\tilde{E}_{y}^{2}+\kappa _{zz}\tilde{E}%
_{z}^{2}+\mu _{zz}\tilde{H}_{z}^{2}+\mu _{yy}\tilde{H}_{y}^{2}\big\}
\label{shpur}
\end{equation}
Here 
\begin{equation}
\hat{K}_{0}=-\partial _{\tau }^{2}-V_{F}^{2}\partial _{x}^{2}+\Delta ^{2},
\end{equation}
and 
\begin{equation}
\kappa _{xx}=\frac{1}{8\pi ^{2}}\frac{\omega _{p}^{2}}{\Delta ^{2}}%
,~~~\kappa _{yy,(zz)}=\kappa _{xx}\left( \frac{t_{y,z}}{\epsilon _{F}}%
\right) ^{2}<<\kappa _{xx}~~~,
\end{equation}
\[
\mu _{yy}\sim \left( \frac{V_{F}}{c}\right) ^{2}\kappa _{zz},~~~\mu
_{zz}\sim \left( \frac{V_{F}}{c}\right) ^{2}\kappa _{yy}
\]
are the diagonal components of the tensor of the dielectric and magnetic
susceptibilities, $\omega _{p}=(8e^{2}V_{F}/\Delta ^{2}b_{y}b_{z})^{1/2}$ is
the plasma frequency, $\epsilon _{F}=V_{F}/b_{x}$ is the Fermi energy, and: 
\[
\tilde{E}_{x}=E_{x}-\frac{1}{2V_{F}e}\partial _{\tau }^{2}\varphi -\frac{%
V_{F}}{2e}\partial _{x}^{2}\varphi \newline
\]
\[
\tilde{E}_{y}=E_{y}-\frac{V_{F}}{2e}\partial _{x,y}^{2}\varphi \newline
\]
\begin{equation}
\tilde{E}_{z}=E_{z}-\frac{V_{F}}{2e}\partial _{xz}^{2}\varphi  \\
\end{equation}
\[
\tilde{H}_{y}=H_{y}+\frac{i}{2V_{F}e}\partial _{\tau z}^{2}\varphi \newline
\]
\[
\tilde{H}_{z}=H_{z}+\frac{i}{2V_{F}e}\partial _{\tau y}^{2}\varphi \newline
\]
Finally, we substitute Eqs. (\ref{shpur}) and (\ref{jakob}) into the Eq. (%
\ref{statsum}) and calculating the variation of the free energy functional
with respect to the vector and scalar potentials to obtain currents in
accordance with Eq.(5). We get in real time: 
\begin{equation}
j_{x}=-\frac{e}{\pi b_{y}b_{z}}\partial _{t}\varphi -\frac{\kappa _{xx}V_{F}%
}{2e}\partial _{t}(\partial _{x}^{2}\varphi -\frac{1}{V_{F}^{2}}\partial
_{t}^{2}\varphi -\frac{V_{y}^{2}}{V_{F}^{2}}\partial _{y}^{2}\varphi -\frac{%
V_{z}^{2}}{V_{F}^{2}}\partial _{z}^{2}\varphi )+\frac{\kappa _{yy}}{2e}%
V_{F}\partial _{yyt}^{3}\varphi +\frac{\kappa _{zz}}{2e}V_{F}\partial
_{zzt}^{3}\varphi ,  \label{jx} \\
\end{equation}
\begin{equation}
\rho =\frac{e}{\pi b_{y}b_{z}}\partial _{x}\varphi -\frac{\kappa _{xx}V_{F}}{%
2e}\partial _{x}(\partial _{x}^{2}\varphi -\frac{1}{V_{F}^{2}}\partial
_{t}^{2}\varphi -\frac{V_{y}^{2}}{V_{F}^{2}}\partial _{y}^{2}\varphi -\frac{%
V_{z}^{2}}{V_{F}^{2}}\partial _{z}^{2}\varphi )+\frac{\kappa _{yy}}{2e}%
V_{F}\partial _{xyy}^{3}\varphi +\frac{\kappa _{zz}}{2e}V_{F}\partial
_{xzz}^{3}\varphi ,  \label{rhoo} \\
\end{equation}
\begin{equation}
j_{y}=-\frac{\kappa _{yy}}{e}V_{F}\partial _{xyt}^{3}\varphi ,  \label{jy} \\
\end{equation}
\begin{equation}
j_{z}=-\frac{\kappa _{zz}}{e}V_{F}\partial _{xzt}^{3}\varphi .  \label{jz}
\end{equation}
It is evident that Eqs. (\ref{jx}) - (\ref{jz}) automatically satisfy the
condition $\partial _{t}\rho +div\vec{j}=0$.

It is easy to check from the equation of motion: 
\begin{equation}
\frac{\delta }{\delta \varphi }\Big\{\int d\vec{r}dt[{\cal L}%
^{ph}+l\{\varphi \}]+Tr~\ln \tilde{L}\Big\}=0  \label{eqmotion1}
\end{equation}
that the polarization correction to Eq. (\ref{eqmotion1}), as well as to $%
j_{x}$ (\ref{jx}) and $\rho $ (\ref{rhoo}), can be safely neglected in
electric fields $|\vec{E}|<<e^{2}/4\pi \kappa _{xx}\Delta ^{2}$. As actually 
$\kappa _{xx}\simeq 1$, this inequality always holds in experiment.

Hence, the incommensurate Q1D CDW is described by the equation of motion: 
\begin{equation}  \label{eqmotion}
\frac{1}{\alpha^2}\partial^2_t\varphi-\frac{V_F^2}{2}\partial^2_x\varphi- 
\frac{U^2_y}{2}\partial^2_y\varphi-\frac{U^2_z}{2}\partial^2_z\varphi= \frac{%
e}{\pi}E_x
\end{equation}
by the Maxwell equations and by the generalized Fr\"ohlich relations: 
\[
\label{jx1} j_x=-\frac{e}{\pi b_yb_z}\partial_t\varphi- +\frac{\kappa_{yy}}{%
2e}V_F\partial^3_{yyt}\varphi+ \frac{\kappa_{zz}}{2e}V_F\partial^3_{zzt}%
\varphi , 
\]
\begin{equation}  \label{rhoo1}
\rho=\frac{e}{\pi b_yb_z}\partial_x\varphi- +\frac{\kappa_{yy}}{2e}%
V_F\partial^3_{xyy}\varphi+ \frac{\kappa_{zz}}{2e}V_F\partial^3_{xzz}\varphi
,
\end{equation}
\[
\label{jy1} j_y=-\frac{\kappa_{yy}}{ e}V_F\partial^3_{xyt}\varphi , 
\]
\[
\label{jz1} j_z=-\frac{\kappa_{zz}}{ e}V_F\partial^3_{xzt}\varphi . 
\]

As an application of our formulas (\ref{eqmotion}), (\ref{rhoo1}), we
calculate the Hall constant of the CDW condensate at zero temperature.

Consider the Hall geometry: $H=H_z$, the transport CDW current $I_x$ flows
parallel to the chains in a film with sizes $L_y,L_z$. As normally $t_z<<t_y$%
, we neglect $z$-dependence in Eqs.(\ref{eqmotion}),(\ref{rhoo1}). Perfoming
the perturbation expansion over $\kappa_{yy}$ in Maxwell equations, we get: 
\begin{equation}
E_y=-2\pi\kappa_{yy}\frac{V_F}{e}\partial^2_{xy}\varphi,~~H_z=H_0
\end{equation}
where $H_0$ is the external magnetic field which is assumed to be larger
than the magnetic field of the transport current.

In sliding CDW $\varphi=\varphi(x-V_Dt,y,z)$, where $V_D(E_x)$ is the
nonlinear drift velocity. We get: 
\begin{equation}
E_y=\frac{2\pi\kappa_{yy}}{e^2}\frac{V_F}{V_D}b_yb_z\frac{\partial j_x} {%
\partial y}
\end{equation}
The Hall constant is: 
\begin{equation}
R_{CDW}=\frac{\int \limits^{L_y}_{0}dy E_y}{cH_zI_x}= \frac{2\pi^2\kappa_{yy}%
}{ce^2H_0}\frac{V_F}{V_D(E_x)}\frac{b_yb_z}{L_z}
\end{equation}
Hence 
\begin{equation}  \label{result}
R_{CDW}I_x(E_x)=const\sim t_y^2 \sim T^2_C
\end{equation}
where $T_C$ is the temperature of the Peierls transition. The r.h.s in Eq.(%
\ref{result}) is independent of $E_x$ which is in a good accordance with the
experimental data \cite{l3}.

{\bf Conclusion.} We have derived for the first time the generalized
Fr\"ohlich relations which relate the transverse currents and fields to the
phase gradients in a Q1D CDW-conductor. The explanation of the relation
between the CDW Hall constant and the nonlinear transport current is found.

$~$ $~$ $~$

\end{document}